# GPU-accelerated finite-temperature Lanczos method for Heisenberg spin systems with non-Abelian permutation symmetries

Shadan Ghassemi Tabrizi[1,2*] and Thomas D. Kühne,[1,2]

[1]*Computational System Sciences, Technische Universität Dresden, 01187 Dresden, Germany*

[2]*Center for Advanced Systems Understanding (CASUS),*
*Am Untermarkt 20, 02826 Görlitz, Germany, *s.ghassemi-tabrizi@hzdr.de*

**Abstract.** We extend a recent GPU implementation of the finite-temperature Lanczos method (FTLM) for Heisenberg spin Hamiltonians, which uses only total-magnetization symmetry, to the full permutation symmetry, including non-Abelian point and space groups. For highly symmetric clusters such as icosahedral polyhedra, the multidimensional irreducible representations (irreps) of the permutation group reduce the block dimensions substantially further than any Abelian subgroup. Here, we provide a matrix-free formalism (avoiding explicit storage of the projected Hamiltonian) for using such symmetries. The implementation applies equally to Abelian groups, retaining the advantages of GPU acceleration. Several cluster topologies are built in; beyond these, symmetry groups can be supplied as site-permutation generators, with irrep matrices computed automatically, so that user-defined clusters run without any code changes. We demonstrate the approach in production runs on a single NVIDIA B200 accelerator for the $s = 3/2$ Heisenberg antiferromagnet on the dodecahedron (Hilbert-space dimension $4^{20} \approx 1.1 \times 10^{12}$) and for the $s = 1/2$ square lattice of $5 \times 8$ sites ($M = 0$ dimension $1.4 \times 10^{11}$). With $R = 24$ random vectors per symmetry block and 60 Lanczos steps, the dodecahedron campaign (with the largest iterated symmetry block having a dimension of $3.6 \times 10^9$) requires about 50 GPU-hours. The code is openly available under the Apache-2.0 license at https://github.com/ghasdeke/ftlm-pg-gpu, archived at DOI: 10.5281/zenodo.21445872.

# 1. Introduction

Finite-temperature properties such as magnetic susceptibility and heat capacity are key probes of exchange-coupled spin clusters and finite fragments of frustrated lattices. When the Hilbert space is too large for full exact diagonalization (ED), the finite-temperature Lanczos method (FTLM) [1,2] combines stochastic trace estimation over random vectors with Krylov-space approximations to thermal expectation values and yields accurate observables at a small fraction of the cost of full diagonalization. Closely related typicality-based approaches obtain thermodynamic observables from one or a few thermal pure quantum states [3–5], and the kernel polynomial method provides an alternative spectral-density route to thermodynamic averages [6]; the accuracy of stochastic estimators is well characterized [7,8]. For the frustrated geometries of interest here, quantum Monte Carlo is severely hampered by the sign problem [9], leaving trace-estimator methods as the tool of choice at finite temperature.

Exploiting spatial symmetries is attractive for FTLM because the Hamiltonian block-diagonalizes over the symmetry sectors. The Lanczos vectors, which dominate the memory footprint, shrink by roughly the order of the group, so larger systems fit into device memory, and each sparse matrix-vector product (SpMV) operates on correspondingly smaller blocks. Of the two structural costs of FTLM – the preparation of the basis data and the repeated SpMV – the former is straightforward without spatial symmetry but demanding to implement efficiently for a non-Abelian group. Systems that benefit from such a treatment include the icosahedron, the dodecahedron, and the icosidodecahedron, which carry the icosahedral point group $I_h$ with irrep dimensions up to five, as well as the cuboctahedron with the octahedral group $O_h$. Such clusters are of long-standing interest in molecular magnetism and frustrated quantum magnetism [10]: Keplerate molecules realize the icosidodecahedron with $s = 5/2$ ($\{Mo_{72}Fe_{30}\}$) [11], $s = 3/2$ ($\{Mo_{72}Cr_{30}\}$) [12], and $s = 1/2$ ($\{Mo_{72}V_{30}\}$) [13], and frustrated high-symmetry polyhedra serve as finite models for extended frustrated lattices such as the kagome antiferromagnet [14–16]. The $s = 1/2$ icosidodecahedron ($N = 30$ sites, $M = 0$ sector dimension $1.6 \times 10^8$) has received extensive theoretical attention [16–18]. Our implementation of Ref. [19] handles systems at this scale on a workstation GPU, but without any point-group reduction. The present work provides the symmetry-adapted treatment of such systems within the same framework.

Combining a magnetization sector ($S^z$ eigenvalue $M$) with a non-Abelian point group is not in itself new. The projection-operator method (see Eq. (9) below) was established for spin clusters by Waldmann [20], and Konstantinidis [21] applied $S^z \otimes I_h$ adaptation, including the multi-dimensional irreps, to the icosahedron and the dodecahedron. The decisive difference is operational: those works (also including the study of Raghu et al. [22]) assemble and store the symmetry-adapted Hamiltonian, so their reach is limited by the memory of this sparse matrix. Large-scale matrix-free ED, on the other hand (the sublattice-coding framework of Wietek and Läuchli [23] and packages such as QuSpin [24] and XDiag [25]) implements symmetry-adapted bases through one-dimensional characters, for which the matrix element collapses to a single phase; a general matrix-element construction for multidimensional irreps is not provided there. Our own GPU-FTLM implementation [19] is matrix-free but uses only $S^z$ symmetry. The gap we close here is a matrix-free, GPU-tractable evaluation of the Hamiltonian action in a basis adapted to non-Abelian permutation symmetry.

GPU implementations are established for ED of itinerant-electron models [26,27] and for wavefunction methods in quantum chemistry [28–30], and mature CPU-based ED frameworks offer finite-temperature capabilities [24,25,31]; performance guidance for sparse matrix-vector products on GPUs is likewise available [32]. We recently showed [19] that a workstation GPU accelerates FTLM by roughly an order of magnitude over an optimized multicore CPU implementation, and that FP32 arithmetic in the SpMV changes thermodynamic observables by amounts far below the statistical uncertainty of the trace estimator. With the block-dimension reduction provided by permutation symmetry and with the larger memory and higher bandwidth of a compute-grade GPU (NVIDIA B200), our present single-GPU implementation completes calculations of a scale previously associated with large CPU clusters. For orientation, ground-state Lanczos ED on distributed-memory machines reached $N = 40$ spin-1/2 sites in 2010 [33] and has since been extended to lattices of up to $N = 50$ [23]; the largest all-sector FTLM thermodynamics computation to date, to our knowledge, is the $N = 42$ kagome lattice [34], obtained with spinpack [35]. The present implementation comes close to this scale (Figure 1) at comparatively modest computational cost.

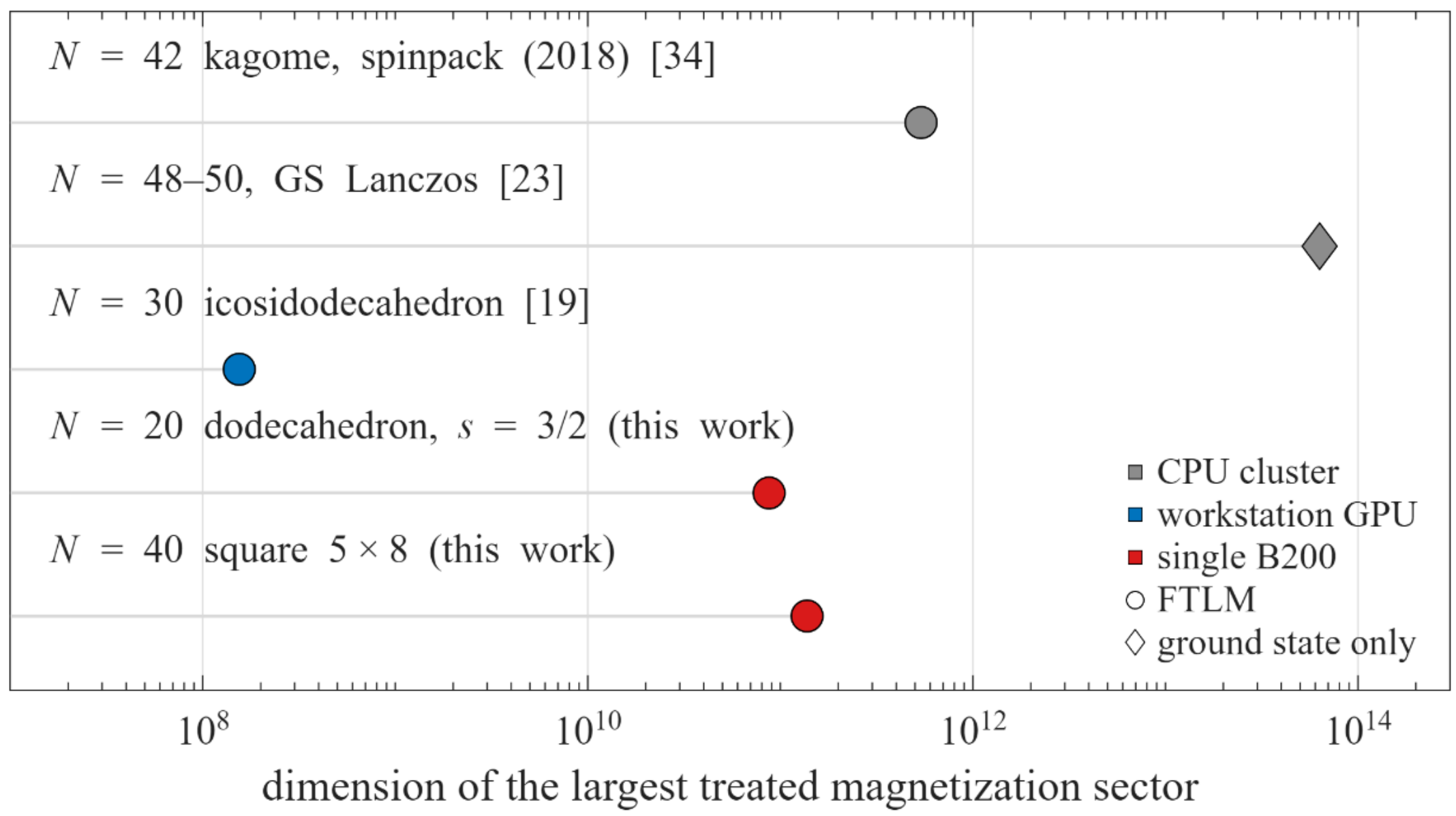

Figure 1: Dimension of the largest treated magnetization sector in representative large-scale spin-system calculations: distributed-memory CPU installations, a workstation GPU, and the single-accelerator runs of this work. Circles denote FTLM calculations, diamonds ground-state-only (GS) Lanczos.

Our contributions are: (i) a closed-form, state-local matrix-element formula (Eq. (15)) that makes the Hamiltonian action matrix-free for multidimensional irreps, obtained from a per-representative eigenvector parametrization of the projector (Section 2.3); (ii) a GPU kernel design that parallelizes over individual (representative, output-row) pairs, so that the SpMV remains efficient for all irrep dimensions (Section 3); (iii) a two-phase construction of the auxiliary data in which the expensive, irrep-independent part is computed only once per magnetization sector and reused for all irreps of the group (Section 3); and (iv) a front-end accepting arbitrary user-supplied permutation groups, whose generators are closed into the full group and whose irrep matrices are computed automatically and brought to real form where possible (Section 3).

The paper is organized as follows: Section 2 sets out the theoretical framework, Section 3 the algorithmic design and GPU implementation, Section 4 verification, benchmarks and reachable system sizes, and Section 5 a summary and outlook.

## 2. Theoretical framework

**2.1 Model and basis encoding.** We consider the Heisenberg model on a cluster of $N$ spin-$s$ sites, Eq. (1),

$$H = J\sum_{\langle i,j\rangle} \mathbf{s}_i \cdot \mathbf{s}_j = J\sum_{\langle i,j\rangle} \left[s_i^z s_j^z + 1/2\left(s_i^+ s_j^- + s_i^- s_j^+\right)\right], \qquad (1)$$

where the sum runs over the $N_B$ bonds of the cluster graph and $J > 0$ corresponds to antiferromagnetic coupling ($J = 1$ throughout Section 4). We work in the product basis $|m_0, \dots, m_{N-1}\rangle$ with $m_i \in \{-s, \dots, s\}$. Since $[H, S^z] = 0$, the Hilbert space decomposes into sectors of total magnetization $M = \sum_i m_i$; explicit computations can be restricted to $M \geq 0$. As in Ref. [19], each basis state is encoded by an integer,

$$n = \sum_{k=0}^{N-1} (m_k + s)\, d^k\,, \qquad (2)$$

where $d = 2s + 1$, so that the shifted local quantum number $a_k = m_k + s$ is the $k$-th base-$d$ digit of $n$. The diagonal action of a bond term is $J\, m_i(n)\, m_j(n)$, and the off-diagonal spin flip generated by $s_i^\pm s_j^\mp$ carries the coefficient of Eq. (3):

$$c_{ij}^{\pm} = \frac{J}{2}\sqrt{s(s+1) - m_i(m_i \pm 1)}\sqrt{s(s+1) - m_j(m_j \mp 1)}. \qquad (3)$$

The $M$-sector basis is the set of labels $n$ whose digits sum to $A_{\text{tot}} = Ns + M$, and its dimension $\mathcal{D}_M$ follows from the counting recursion of Ref. [36].

**2.2 FTLM with symmetry blocks.** For a Hilbert-space block of dimension $\mathcal{D}$, a set of $R$ normalized random vectors $|r\rangle$ gives the stochastic trace estimate of Eq. (4) for the partition function $Z$,

$$Z \approx \frac{\mathcal{D}}{R}\sum_{r=1}^{R} \langle r|e^{-\beta H}|r\rangle \quad , \qquad (4)$$

where $\beta = 1/k_B T$. Each matrix element is approximated in a Krylov space. $N_L$ Lanczos steps started from $|r\rangle$ yield Ritz values $\theta_k$ and spectral weights $w_k$ (the squared first components of the

eigenvectors of the tridiagonal matrix of each chain), satisfying the sum rule $\sum_k w_k = 1$. When $H$ commutes with a spatial symmetry group $G$, the estimator factorizes over $(M, \Gamma)$ blocks, where $\Gamma$ labels the irreps of $G$. Working at a single partner row (irrep component), each block enters with multiplicity $d_\Gamma$, Eq. (5),

$$Z_{\text{FTLM}} \approx \sum_{M \geq 0} \text{mult}_M \sum_\Gamma d_\Gamma \frac{\mathcal{D}_{M,\Gamma}}{R} \sum_{r=1}^{R} \sum_{k=1}^{N_L} w_k^{(M,\Gamma,r)} e^{-\beta\theta_k^{(M,\Gamma,r)}}, \qquad (5)$$

with $\text{mult}_M = 2$ for $M > 0$ (sectors with $\pm M$ contribute equally) and $\text{mult}_M = 1$ for $M = 0$, and $\mathcal{D}_{M,\Gamma}$ the block dimension. The random vectors are drawn independently per block: real normalized Gaussian vectors in real-arithmetic blocks (Section 3), complex ones otherwise; very small blocks can alternatively be diagonalized directly. In the $M = 0$ sector the Hamiltonian additionally commutes with the global spin flip, $S^z \to -S^z$, which classifies the levels by a parity $(-1)^S$ determined by the total spin $S$. We denote the corresponding two-element group by $Z_2$. Extending $G$ to $G \times Z_2$ splits each irrep into a pair $\Gamma^\pm$ (in the $M = 0$ sector, $\Gamma$ henceforth labels irreps of this extended group), which roughly halves every block dimension at negligible extra cost (the flip image of a state is available arithmetically, so the search for orbit representatives, introduced in Section 2.3, requires no additional permutation operations).

Thermal energy moments ($p = 1,2$) follow analogously,

$$\langle H^p \rangle_{\text{FTLM}} \approx \frac{1}{Z_{\text{FTLM}}} \sum_{M \geq 0} \text{mult}_M \sum_\Gamma d_\Gamma \frac{\mathcal{D}_{M,\Gamma}}{R} \sum_{r,k} w_k^{(M,\Gamma,r)} \left[\theta_k^{(M,\Gamma,r)}\right]^p e^{-\beta\theta_k^{(M,\Gamma,r)}} \qquad (6)$$

giving the heat capacity $C = \beta^2(\langle H^2 \rangle - \langle H \rangle^2)$. The zero-field susceptibility is evaluated from the sector partition functions $Z_M = \sum_\Gamma d_\Gamma Z_{M,\Gamma}$, where $Z_{M,\Gamma} = \frac{\mathcal{D}_{M,\Gamma}}{R} \sum_{r,k} w_k^{(M,\Gamma,r)} e^{-\beta\theta_k^{(M,\Gamma,r)}}$ is the per-block partition-function estimate, so that Eq. (5) reads $Z_{\text{FTLM}} \approx \sum_{M \geq 0} \text{mult}_M Z_M$ (units $k_B = 1$, $\mu_B = 1$; unless stated otherwise, $\chi$ is quoted for $g = 1$):

$$\chi = g^2 \beta \langle (S^z)^2 \rangle = g^2 \beta \frac{\sum_{M \geq 0} \text{mult}_M M^2 Z_M}{\sum_{M \geq 0} \text{mult}_M Z_M}. \qquad (7)$$

Eq. (7) omits the $\langle S^z \rangle^2$ term of the general fluctuation formula, because $\langle S^z \rangle = 0$ at zero field.

**2.3 Symmetry-adapted basis.** This section constructs an orthonormal basis of each $(M, \Gamma)$ block, together with the per-orbit quantities entering the matrix elements derived in Section 2.4.

Let $G$ be a finite group of lattice symmetries acting on integer-label states (Eq. (2)) by site permutations. For $g \in G$ with permutation $\pi_g$, the digits transform as $a_k \mapsto a_{\pi_g(k)}$. The orbit of $n$ under $G$ is $O(n) = \{U(g)|n\rangle: g \in G\}$; as the orbit representative we choose the state whose label is smallest, $r \equiv \min O(n)$. The stabilizer subgroup of the representative is given in Eq. (8):

$$\mathrm{Stab}(r) = \{g \in G: U(g)|r\rangle = |r\rangle\} \quad . \quad (8)$$

For each irrep $\Gamma$ of $G$ with dimension $d_\Gamma$, character $\chi_\Gamma$, and representation matrices $\boldsymbol{\rho}_\Gamma(g)$ (from standard tables [37] or computed numerically, Section 3), the projector onto the irrep block within the orbit-spanned subspace is given in Eq. (9),

$$\Pi_\Gamma^{(r)} = \frac{d_\Gamma}{|G|} \sum_{g \in G} \chi_\Gamma(g)^* \, U(g) \quad . \quad (9)$$

Before projected states can serve as basis vectors, one must determine how many linearly independent states the projector produces from a given orbit, and their mutual overlaps. Both are contained in the Gram matrix of the row-projected states $P^\Gamma_{\alpha,\beta}|r\rangle$, $\beta = 1, \dots, d_\Gamma$, taken at any fixed partner row $\alpha$ (the result is $\alpha$-independent), where $P^\Gamma_{\alpha,\beta} = \frac{d_\Gamma}{|G|} \sum_{g \in G} [\boldsymbol{\rho}_\Gamma(g)^*]_{\alpha,\beta} \, U(g)$ is a transfer operator, which resolves the character projector of Eq. (9) into individual partner rows. Because all states in the orbit are permutations of $|r\rangle$, this Gram matrix reduces to a sum of conjugated representation matrices over the stabilizer alone, Eq. (10):

$$\left(\mathbf{M}_\Gamma^{(r)}\right)_{\beta,\beta'} = \frac{d_\Gamma}{|G|} \sum_{h \in \mathrm{Stab}(r)} [\boldsymbol{\rho}_\Gamma(h)^*]_{\beta,\beta'} \quad . \quad (10)$$

This $d_\Gamma \times d_\Gamma$ matrix is the only orbit-specific group-theoretic object required here; for orbits of full length it is $\left(\frac{d_\Gamma}{|G|}\right)$ times the identity, so $n_\Gamma(r) = d_\Gamma$. Since $\left(\frac{1}{|\mathrm{Stab}(r)|}\right) \sum_h \boldsymbol{\rho}_\Gamma(h)^*$ is an orthogonal projector, all non-zero eigenvalues of $\mathbf{M}_\Gamma^{(r)}$ equal $d_\Gamma |\mathrm{Stab}(r)|/|G|$, and their count is $n_\Gamma(r) = \left(\frac{1}{|\mathrm{Stab}(r)|}\right) \sum_h \chi_\Gamma(h)^*$. Diagonalization, Eq. (11),

$$\mathbf{M}_\Gamma^{(r)} = \mathbf{V}_r \, \boldsymbol{\Lambda}_r \, \mathbf{V}_r^\dagger, \qquad \boldsymbol{\Lambda}_r = \mathrm{diag}\left(\lambda_{r,1}, \dots, \lambda_{r,n_\Gamma(r)}, 0, \dots, 0\right), \qquad (11)$$

provides this information: the number $n_\Gamma(r)$ of strictly positive eigenvalues is the number of independent symmetry-adapted basis vectors (at the fixed partner row) contributed by the orbit of $r$ to irrep $\Gamma$ (zero if the orbit is incompatible with $\Gamma$), the eigenvalues $\lambda_{r,k}$ provide normalizations, and the eigenvectors collected in $\mathbf{V}_r$ specify how the stabilizer directions combine. Because all non-zero eigenvalues are equal, the eigenvalue ratio entering Eq. (15) below reduces to $\sqrt{\frac{\lambda_{r\prime,k\prime}}{\lambda_{r,k}}} = \sqrt{\frac{|\mathrm{Stab}(r\prime)|}{|\mathrm{Stab}(r)|}} = \sqrt{\frac{L_r}{L_{r\prime}}}$ in terms of the orbit lengths $L_r = |G|/|\mathrm{Stab}(r)|$, which is the orbit-size information used in Section 2.4. Orbit representatives with $n_\Gamma(r) > 0$ are the active representatives of $\Gamma$. The symmetry-adapted basis vectors of Eq. (12), with $k = 1, \dots, n_\Gamma(r)$ and $\alpha = 1, \dots, d_\Gamma$, are orthonormal across $(r, k, \alpha, \Gamma)$ because of the orthogonality relations and Eq. (11); they block-diagonalize $H$ by $(M, \Gamma, \alpha)$.

$$|\psi_{r,k,\alpha,\Gamma}\rangle = \frac{1}{\sqrt{\lambda_{r,k}}} \sum_{\beta=1}^{d_\Gamma} (\mathbf{V}_r)_{\beta,k} P^\Gamma_{\alpha,\beta} |r\rangle = \frac{d_\Gamma}{|G|\sqrt{\lambda_{r,k}}} \sum_{g \in G} [\boldsymbol{\rho}_\Gamma(g)^* \, \mathbf{V}_r]_{\alpha,k} \, U(g)|r\rangle \qquad (12)$$

In the one-dimensional limit, $\mathbf{M}_\Gamma^{(r)}$ reduces to the squared norm of the symmetry-projected state, and $n_\Gamma(r) \in \{0,1\}$ indicates whether that state is nonzero.

Two structural properties reduce the computational effort. First, the per-$\alpha$ blocks of $H$ are spectrally identical, i.e., $H$ commutes with every $U(g)$ and hence with the transfer operators resolving the $(M, \Gamma)$ subspace into partner rows; by Schur's lemma $H$ acts identically on each row, so the spectrum is the same for every $\alpha$. We therefore build and iterate only the $\alpha = 1$ block, whose dimension (entering Eq. (5)) is

$$\mathcal{D}_{M,\Gamma} = \sum_{r \in \text{active reps}(\Gamma)} n_\Gamma(r), \qquad (13)$$

and the trace aggregation includes the $\alpha > 1$ blocks through the multiplicity factor,

$$w_k \leftarrow w_k \cdot \mathrm{mult}_M \cdot d_\Gamma \quad , \qquad (14)$$

which is the only place where $d_\Gamma$ enters the trace evaluation directly. The factor is applied once, at aggregation time, so the per-chain sum rule remains intact. The $\alpha \neq 1$ copies are never constructed, and there is no per-$\alpha$ overhead in the SpMV or in memory. Eq. (14) implements the factors $\text{mult}_M \cdot d_\Gamma$ appearing in Eqs. (5) and (6), which are therefore not applied again. Second, all group sums have been absorbed into the per-orbit data $(\lambda_{r,k}, \mathbf{V}_r)$ during the construction of the basis, and no group summation remains in the Hamiltonian application (Section 2.4). For the $I_h$ group the advantage over an Abelian-subgroup treatment is quantitative: the largest Abelian subgroup has order 10, capping the Abelian block reduction at 10×, whereas a single partner row of the full group leaves a total iterated dimension of only $\sum_\Gamma \mathcal{D}_{M,\Gamma} \approx \left(\frac{\sum_\Gamma d_\Gamma}{\sum_\Gamma d_\Gamma^2}\right) \mathcal{D}_M = (32/120)\mathcal{D}_M \approx 0.27\mathcal{D}_M$ and reduces the largest block dimension, which determines the memory required for the Lanczos vectors, from $\mathcal{D}_M/10$ to about $\mathcal{D}_M/24$ (Figure 2).

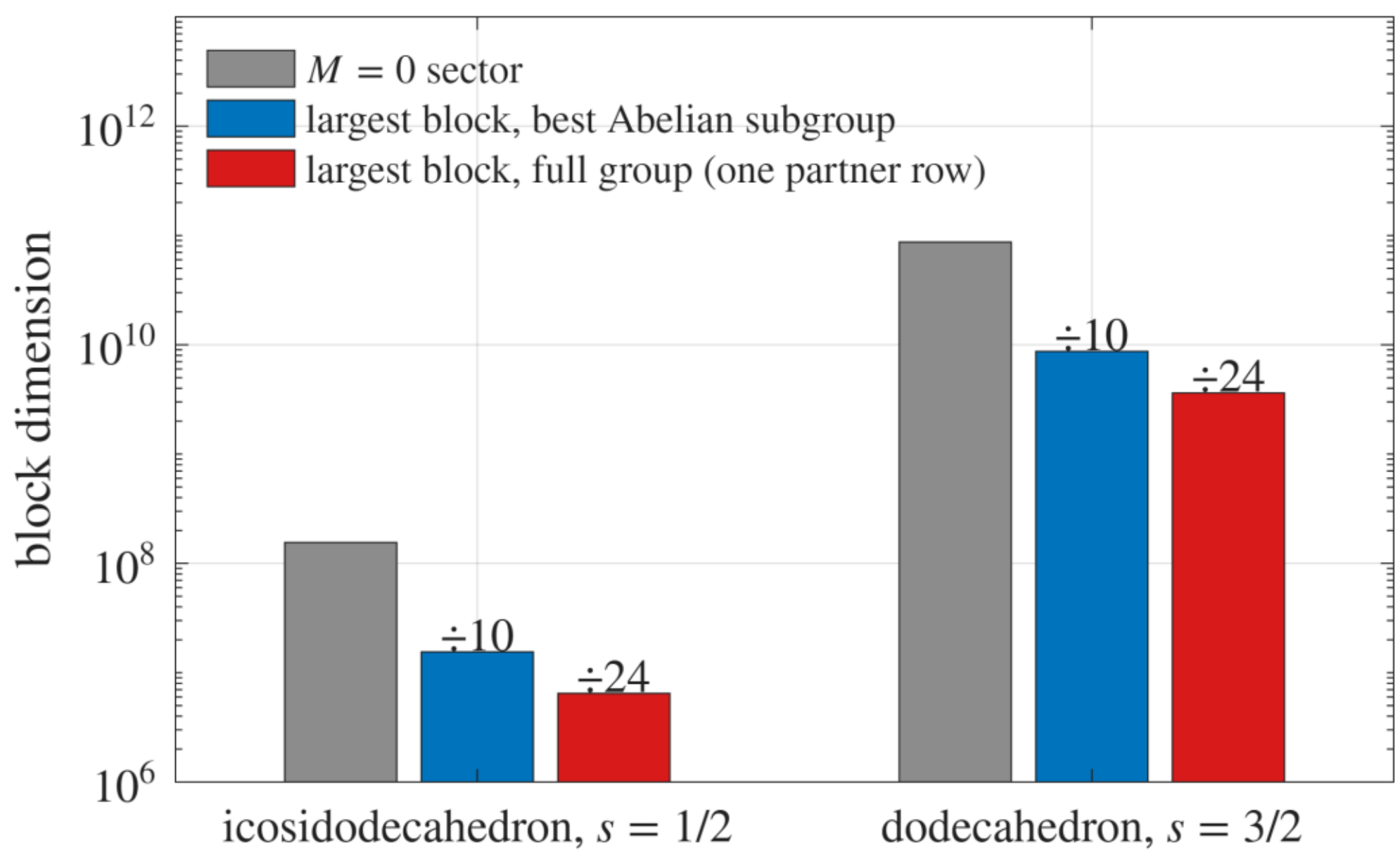

Figure 2: Block-size reduction by the full non-Abelian symmetry: for the $M = 0$ sectors of the $s = 1/2$ icosidodecahedron and the $s = 3/2$ dodecahedron, the sector dimension is compared with the largest block under the best Abelian subgroup of $I_h$ (order 10) and with the largest single-partner-row block of the full group (Eq. (13)). For the dodecahedron the corresponding FP32 Lanczos vectors take 348 GB, 35 GB, and 14.5 GB, respectively.

**2.4 Matrix-element formula.** Applying the off-diagonal (ladder) terms of $H$ to a basis vector $|\psi_{r,k,1,\Gamma}\rangle$ produces, for each bond and each admissible sign choice in Eq. (3), a state $c_a|a\rangle$, where $a$ is the integer label (Eq. (2)) of the flipped state and $c_a$ the coefficient of Eq. (3). Decomposing $a$ into its orbit minimum, $|a\rangle = U(g_e)|r'\rangle$ with $r' = \min O(a)$, and substituting into Eq. (12), the off-diagonal matrix element in the symmetry-adapted basis takes the compact closed form

$$M^{(e)}_{k',k} = c_a \sqrt{\frac{\lambda_{r',k'}}{\lambda_{r,k}}} \left[\mathbf{V}^{\dagger}_{r'}\, \boldsymbol{\rho}_{\Gamma}(g_e)^T\, \mathbf{V}_r\right]_{k',k} \quad . \qquad (15)$$

The element $g_e$ is unique only up to the stabilizer of $r'$. Eq. (15) is invariant under this freedom because the stabilizer sum is absorbed into $\mathbf{V}_{r'}$. A self-contained derivation of Eq. (15) is given in Appendix A. In practice, $g_e$ is obtained as a by-product of the search for the orbit representative. Two features deserve emphasis: the coupling matrix per pair of representatives is small ($d_\Gamma \times d_\Gamma$) and independent of the specific system with the given symmetry, and the $\boldsymbol{\rho}_\Gamma$ contraction is its only group-element-dependent factor, the eigenvalue ratio having reduced to orbit lengths (Section 2.3). The diagonal contribution is irrep- and stabilizer-invariant,

$$\langle \psi_{r,k,1,\Gamma} | H^{\text{diag}} | \psi_{r,k,1,\Gamma} \rangle = J \sum_{\langle i,j \rangle} m_i\,(r)\, m_j(r) \quad , \qquad (16)$$

evaluated once per active representative, independently of $(k, \Gamma)$. In matrix form the diagonal part is strictly diagonal in both indices, $\langle \psi_{r',k',1,\Gamma} | H^{\text{diag}} | \psi_{r,k,1,\Gamma} \rangle = \delta_{r,r'}\, \delta_{k,k'}$ Eq. (16). Off-diagonal elements vanish because the bond list is invariant under $G$, so the diagonal energy is constant along each orbit.

For one-dimensional irreps, Eq. (15) reduces to the standard construction of momentum-space ED, i.e., the projector contributes a single character phase and a scalar normalization, which is how matrix-free symmetrized ED has operated since the work of Lin [38] and in the frameworks of Refs. [23,24,39]. For $d_\Gamma > 1$, $\mathbf{V}_r$ replaces the scalar normalization and $n_\Gamma(r)$ the binary compatibility test (Section 2.3); every quantity in Eq. (15) remains attached to a single representative or group element, so no global symmetrization matrix and no stored matrix elements are required (Section 3).

Whether the multi-dimensional treatment reduces the block dimensions below what one-dimensional characters can already achieve depends on the structure of the group. For the periodic lattices considered in this work, the symmetry group is a semidirect product $T \rtimes P$ of the Abelian translation group $T$ with the point group $P$. The translations form an Abelian normal subgroup with super-solvable quotient, so every irrep is monomial, i.e., induced from a one-dimensional character of a suitable subgroup [40,41]. By Frobenius reciprocity, the single-partner-row block of a monomial irrep is unitarily equivalent to the subspace obtained by projecting with that one-dimensional character and can therefore be computed with phase factors alone. For $T \rtimes P$ this induction is the established momentum-space procedure: one momentum is selected per point-group star, one-dimensional characters of the little group (or of a suitable subgroup) are applied, and the star size enters as a degeneracy factor. For these lattices, the non-Abelian treatment therefore yields the same block dimensions as momentum-space techniques and does not enlarge the reachable system size, but it organizes the computation uniformly, without a separate implementation path. The polyhedral clusters are different: the rotation group of the icosahedron is isomorphic to the alternating group $A_5$, which is not solvable, whereas monomial groups are necessarily solvable [41]. Since $A_5$ has no subgroup of index smaller than five, the three- and four-dimensional irreps are not induced from one-dimensional characters (the five-dimensional irreps are monomial, induced from a nontrivial one-dimensional character of the index-five subgroup $A_4$), and the reduction quantified in Section 2.3 cannot be reproduced by one-dimensional projections. The octahedral group of the cuboctahedron is likewise monomial – it contains an Abelian normal subgroup of order eight (the three twofold rotations together with the inversion) with super-solvable quotient [40,41] – but no translation structure organizes the corresponding inductions, and the uniform multi-dimensional treatment is the practical route there as well.

# 3. Algorithmic design and GPU implementation

The computation separates into a preprocessing part and the Lanczos iteration. In the former, a list of entries is built for each $(M, \Gamma)$ block, where each entry records a pair of coupled representatives, the connecting group element, and the spin coefficient of Eq. (3). The Lanczos iteration then applies the Hamiltonian repeatedly through the SpMV, which evaluates Eq. (15) for every entry from the small matrices $\mathbf{V}_r$ and $\boldsymbol{\rho}_\Gamma$ and dominates the run time (69–93% in the benchmarks of Section 4). The symmetry-adapted Hamiltonian is thus never assembled or stored; the memory footprint is set by the entry list, the basis data, and the Lanczos vectors. The FP32/FP64 division of labor, the bitmap lookup table, and the GPU-adapted combinatorial ranking are unchanged from Ref. [19] and are only briefly summarized below. A single CUDA kernel serves all cluster geometries. The direct evaluation of the Hamiltonian action is independent of the trace estimator built on top of it, so thermal-pure-quantum-state propagation [3–5] or kernel-polynomial expansions [6] could use the same SpMV kernel.

Each entry corresponds to one admissible pair spin flip (a source representative, a bond, and a sign choice in Eq. (3)). Entries are sorted by the representative of the output row, so that all contributions to one row are contiguous. The construction is split into an irrep-independent and an irrep-dependent part. The former, which is performed only once per magnetization sector, computes the state generated by each bond spin flip, determines the representative of its orbit, records the connecting group element, and evaluates the coefficient of Eq. (3), where the search for the orbit representative is the most expensive step. The irrep-dependent part assembles for each $\Gamma$ the per-representative data and the per-group-element tables entering Eq. (15) by small batched matrix products. Sublattice-coding techniques [23] could accelerate the representative search further but were not pursued here.

The entry construction requires a map from the integer label of the state generated by a bond spin flip (Eq. (3)) to its position in the sorted representative list. The SpMV itself uses precomputed source indices and performs no lookups. Two interchangeable lookup methods, both described in Ref. [19], serve this purpose; the choice between them is set by the range of the state label and affects only the preparation stage. The bitmap-based compressed lookup table is the default for $\mathcal{D}_{\text{full}} \leq 2^{32}$ (its memory grows linearly with $\mathcal{D}_{\text{full}}$, reaching about 1 GB at that limit). The GPU-adapted combinatorial ranking of Schnack et al. [36] (a few kilobytes of tables), here combined with a binary search in the sorted representative rank array, has no such restriction and is used for the largest systems of Section 4 ($\mathcal{D}_{\text{full}} \sim 10^{12}$). In practice the choice does not significantly affect the total wall time.

Entry tables larger than the device memory are kept in host memory and streamed to the GPU during each SpMV in contiguous segments ordered by output row; tables that exceed the host memory as well are assembled and sorted on disk and streamed from there, with disk reads, host-to-device transfers, and kernel execution overlapping. A leading portion of the table, sized to the free device memory, stays resident, so only the remainder is re-transferred. Because a streamed pass costs roughly one traversal of the table per SpMV, the run-time selection gives full table residency precedence over a larger batch width (itself capped at $B = 8$ by the register budget of the kernel). Besides the Lanczos vectors, only the compact per-representative basis data and the resident portion of the table occupy the device permanently.

Each of the $R$ Gaussian random starting vectors carries its own independent three-term Lanczos recurrence. Chains are processed in batches of width $B \leq 8$ that share each SpMV launch; the batch width is chosen at run time from the free device memory, so the same input deck adapts to GPUs of different memory size. The resulting per-chain tridiagonal matrices are diagonalized on the host in FP64 at negligible cost to give the Ritz pairs entering Eqs. (4) and (6). The starting vectors are complex for irreps treated in complex arithmetic and real for irreps in real form (see below). As shown in Ref. [19], the faster loss of orthogonality of the Lanczos recursion in FP32 does not significantly affect thermodynamic observables, independently of system size.

In addition to this FP32 default, the kernel provides an optional half-precision storage mode for the three Lanczos work vectors: every read from memory converts FP16 to FP32 on the fly, all arithmetic (the SpMV accumulation of Eq. (15), the dot products, and the three-term recurrence) is carried out in FP32, and every write rounds to nearest FP16. The recursion coefficients are computed from the vectors as actually stored. Because the SpMV is dominated by random gathers of vector elements, halving the element size halves both the gather traffic and the vector footprint, thereby doubling the feasible batch width $B$. Section 4.2 quantifies the accuracy of this mode.

The SpMV kernel evaluates Eq. (15) on the fly, reconstructing the contraction $V_{r'}^{\dagger}\boldsymbol{\rho}_{\Gamma}(g_e)^T V_r$ for each entry from the small per-representative and per-group tables. In the one-dimensional case this contraction reduces to a single scalar phase; for higher irrep dimensions the kernel parallelizes over pairs of output representative and output row, so that each thread accumulates a single row and the per-thread state stays small (one thread per representative would have to hold all $d_{\Gamma}$ rows across the batch of Lanczos chains, exceeding the register budget for the larger irreps). The kernel operates efficiently for all irrep dimensions supported by the released code ($d_{\Gamma} \leq 12$).

The group enters the computation only through two tables, the site permutations and the irrep matrices. Built-in providers cover the icosahedron, dodecahedron and icosidodecahedron ($I_h$, $|G| = 120$), the cuboctahedron ($O_h$, $|G| = 48$), periodic square fragments of user-specified size $m \times n$ (the translation subgroup alone can also be selected), and a few specific high-symmetry kagome and triangular lattice fragments with their full space groups (point group ⋉ translations). Beyond these, an arbitrary system can be supplied by listing only the generators of its symmetry group as site permutations, together with the bond list: the generators are closed into the full group (Dimino's algorithm [42]), all irreps are computed numerically as explicit unitary matrices from the regular representation, and irreps of real type in the Frobenius–Schur classification (indicator +1) are brought to real orthogonal form by a change of basis. The generated irreps are verified at build time (unitarity of the representation matrices, character orthogonality, and the completeness relation $\sum_{\Gamma} d_{\Gamma}^2 = |G|$) with residuals near machine precision for all built-in groups, and the real form is accepted only when the transformed matrices pass the same tests; the limit $d_{\Gamma} \leq 12$ of the kernel is an implementation constant. Real representation matrices allow the entire downstream computation to run in real arithmetic, which halves the memory demand and memory traffic of the FP32 GPU kernel and, together with the entry streaming described above, brings the $s = 3/2$ dodecahedron within reach of a single B200 GPU. For irreps of complex or quaternionic type (indicator 0 or $-1$), a complex-arithmetic path is retained. Two automatic checks protect against inconsistent user input: the closure step guarantees that the supplied permutations generate a group, and the bond list is verified to be invariant under every group element.

The method is implemented in MATLAB with custom CUDA kernels. Development and verification were performed on a workstation with an NVIDIA RTX 4000 SFF Ada GPU (20 GB of device memory). Production runs for the largest systems used a single NVIDIA B200 accelerator (192 GB), hosted on a node that provides 2.3 TB of RAM and node-local NVMe scratch storage for the out-of-core stages of Section 3. All wall times are end to end, and quoted memory figures are peak values.

# 4. Results

**4.1 Verification.** Figure 3 compares the heat capacity $C(T)$ of the $s = 3/2$ icosahedron and cuboctahedron, computed with the present implementation using combined $S^z$ and full point-group adaptation ($I_h$ and $O_h$, respectively; $R = 100$ random vectors per block, and $N_L = 100$ Lanczos steps), against full ED reference data from Ref. [43], which are independent of the present code. Deviations between the curves are consistent with the statistical uncertainty of the trace estimator at this sampling.

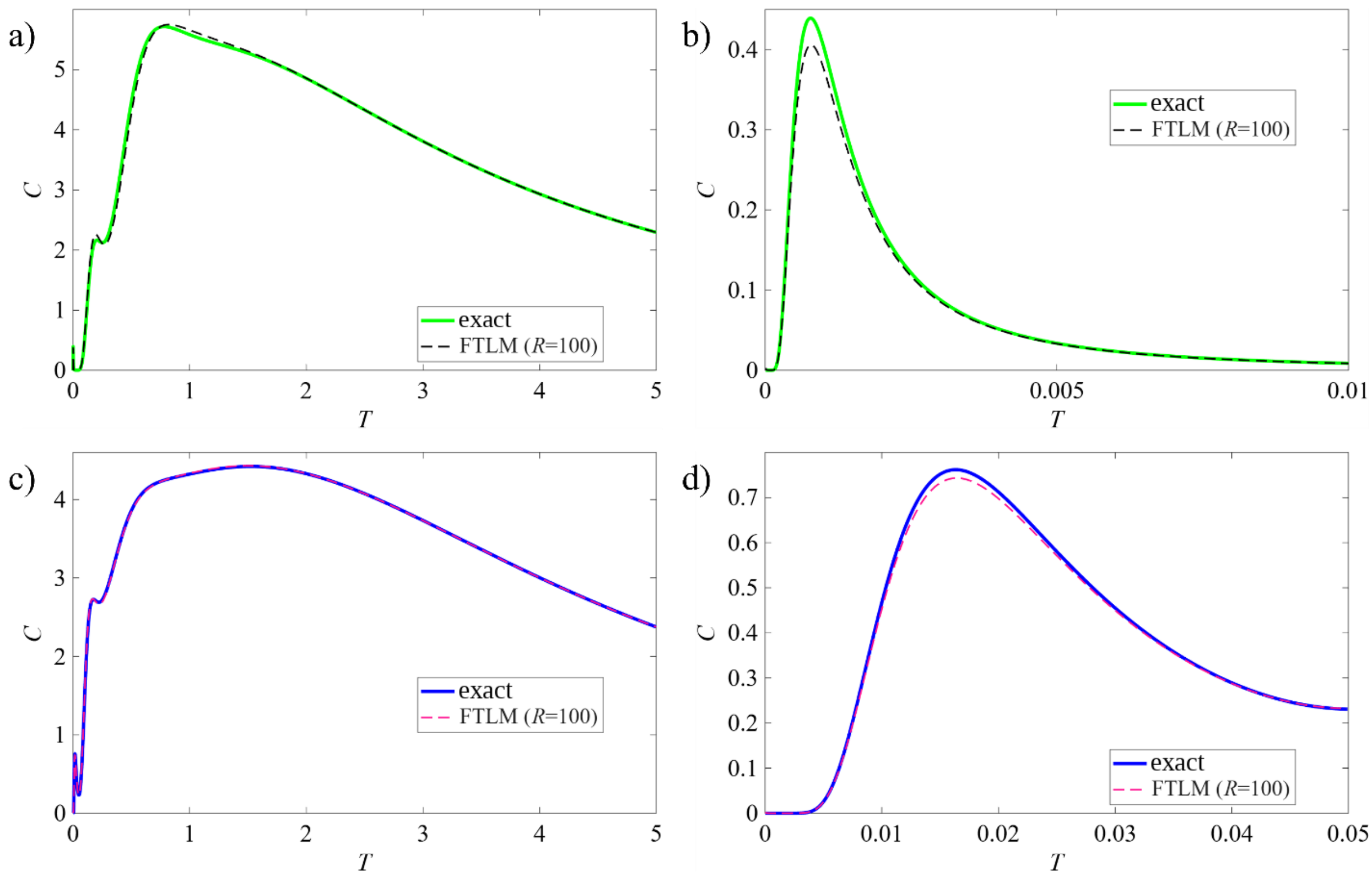

Figure 3: Heat capacity $C(T)$ of the $s = 3/2$ icosahedron (top row: panels (a) and (b)) and cuboctahedron (bottom row: panels (c) and $(d)$), computed by full ED (data from Ref. [43]) and with the present FTLM code. The right-hand panels (b) and $(d)$ resolve the low-temperature peak.

Each symmetry-adapted Hamiltonian block is additionally checked against a dense reference (the projector of Eq. (9) applied to the full $M$-sector Hamiltonian): the spectra agree numerically for all ten $I_h$ irreps in every sector small enough for the dense construction, for the $s = 1/2$ and $s = 3/2$ icosahedron and selected icosidodecahedron blocks; an automated regression suite cross-checks the remaining components (lookup methods, GPU versus CPU orbit search, generated versus tabulated irreps, the memory tiers of Section 3, and the GPU SpMV against a CPU FP64 reference). Finally, two exact high-temperature limits validate the trace aggregation at production scale: for the dodecahedron campaign of Section 4.3, the Curie constant $\lim_{T\to\infty} T\chi/g^2 = Ns(s+1)/3$ is reproduced to machine precision and $\mathrm{Tr}H/D = 0$ to five parts in $10^6$ of the spectral width.

**4.2 Half-precision vector storage.** The FP16 storage mode of Section 3 was validated in the same way as the FP32 arithmetic in Ref. [19]. First, for the $s = 3/2$ icosahedron ($R = 100$, $N_L = 100$), the fully $I_h$ adapted pipeline was run once with FP32 and once with FP16 vector storage (with identical random seeds): for $T \geq 0.1$ the heat capacity and susceptibility change by at most $|\Delta C| = 8 \times 10^{-5}$ and $|\Delta\chi| = 10^{-4}$, roughly three orders of magnitude below the statistical uncertainty of the trace estimator. Second, we ported the same storage scheme to the unsymmetrized implementation of Ref. [19] and recomputed its complete FP32-versus-FP64 precision test set against FP64 CPU references. Across all six systems spanning several orders of magnitude in Hilbert-space dimension, the deviations in $C$ and $\chi$ stay approximately between $10^{-5}$ and $10^{-4}$. Figure 4 shows this comparison for the $s = 3/2$ icosahedron.

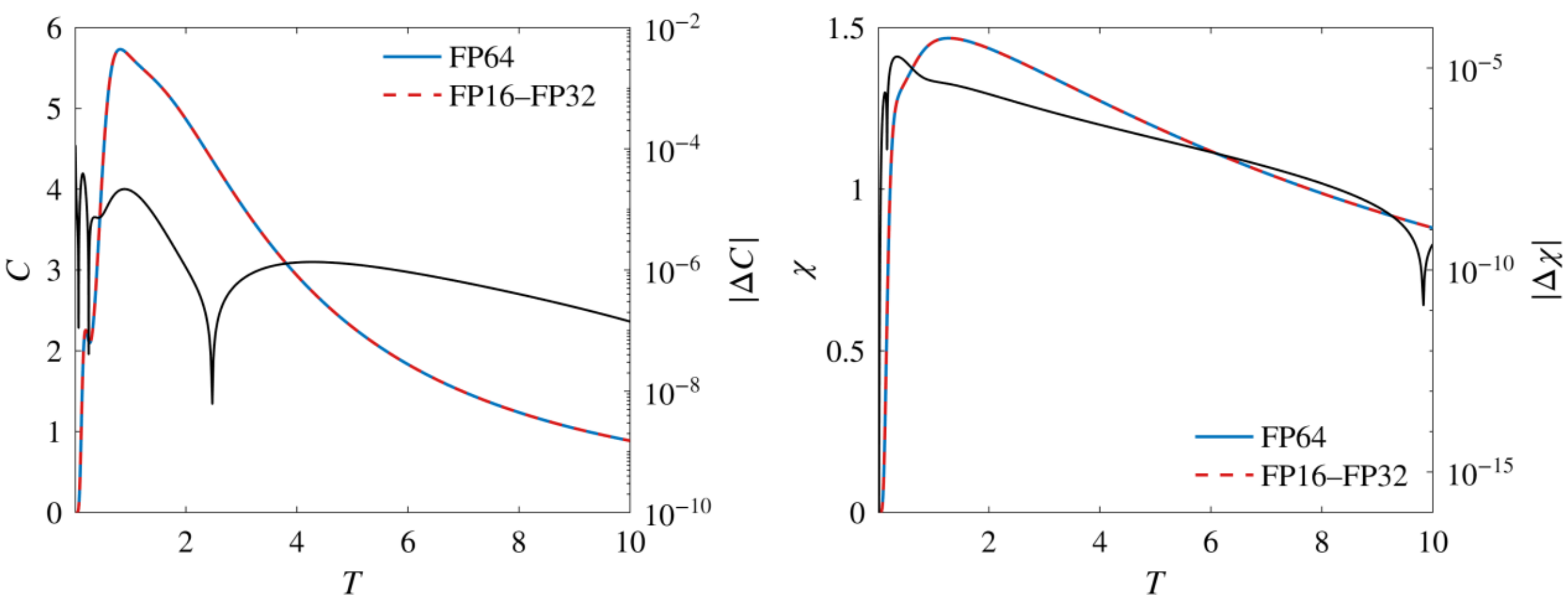

Figure 4: Heat capacity (left) and zero-field susceptibility (right) of the $s = 3/2$ icosahedron, computed with the implementation of Ref. [19] using FP64 vector storage and using the FP16 storage mode of Section 3 (identical random seeds). The curves coincide at the resolution of the plot, and the FP64 reference reproduces the corresponding curves of Ref. [19]. The pointwise deviation, $|\Delta C|$ or $|\Delta\chi|$, is plotted against the logarithmic right axis.

These deviations are consistent with a heuristic backward-error picture [19]. Rounding the stored iterates acts, to leading order, like an operator-level perturbation $\|\delta H\| = \mathcal{O}(u_{\text{eff}}\, W)$ [44], where $W$ is the spectral bandwidth of $H$ and $u_{\text{eff}}$ is the effective roundoff of the storage format (the arithmetic itself remains FP32 and the recursion coefficients are computed from the stored, quantized iterates, so each $\beta$ is the norm of exactly the vector that enters the next step). The measured deviations lie well below this worst-case envelope. For the largest blocks, $u_{\mathrm{eff}}$ itself grows with the block dimension: once the typical component of a normalized Lanczos vector, $n_{\mathrm{basis}}^{-1/2}$, falls below the smallest normal FP16 value of $2^{-14} \approx 6.1 \times 10^{-5}$ (block dimensions $n_{\mathrm{basis}} \gtrsim 3 \times 10^{8}$), the components are stored on the subnormal grid of FP16 with fixed absolute quantum $2^{-24}$, so that $u_{\mathrm{eff}} \approx n_{\mathrm{basis}}^{1/2}\, 2^{-24}$ – about $3.6 \times 10^{-3}$ for the largest campaign blocks instead of the nominal $2^{-11}$. The same-seed FP32 reference of that campaign operates at precisely this scale, so the deviation quoted in Section 4.3 includes the subnormal contribution. Storing each block's vectors with a per-block power-of-two scale factor removes this dimension dependence. It is implemented in the released code; in a same-seed repetition of the $M = 0$ sector it reduces the deviation of the specific-heat trace from the FP32 reference by almost two orders of magnitude, while the ground-state offset remains at the same few-parts-in-$10^{4}$ level, and the subnormal quantum affects the thermodynamic traces, not the extremal Ritz values. The validation is therefore expected to carry over to larger systems.

The same-seed comparison extends to the fully symmetry-adapted pipeline at production settings. For the $s = 3$ icosahedron (all magnetization sectors, batch width pinned, so that the random vectors and hence the Krylov chains are identical and only the storage format differs), the FP16 run changes $C(T)$ and $\chi(T)$ by at most $3 \times 10^{-4}$ of the respective curve maxima and the ground-state estimate by $-5.0 \times 10^{-4}$ (Figure 5).

The following section quantifies the performance gains that the halved vector footprint enables on memory-bound blocks.

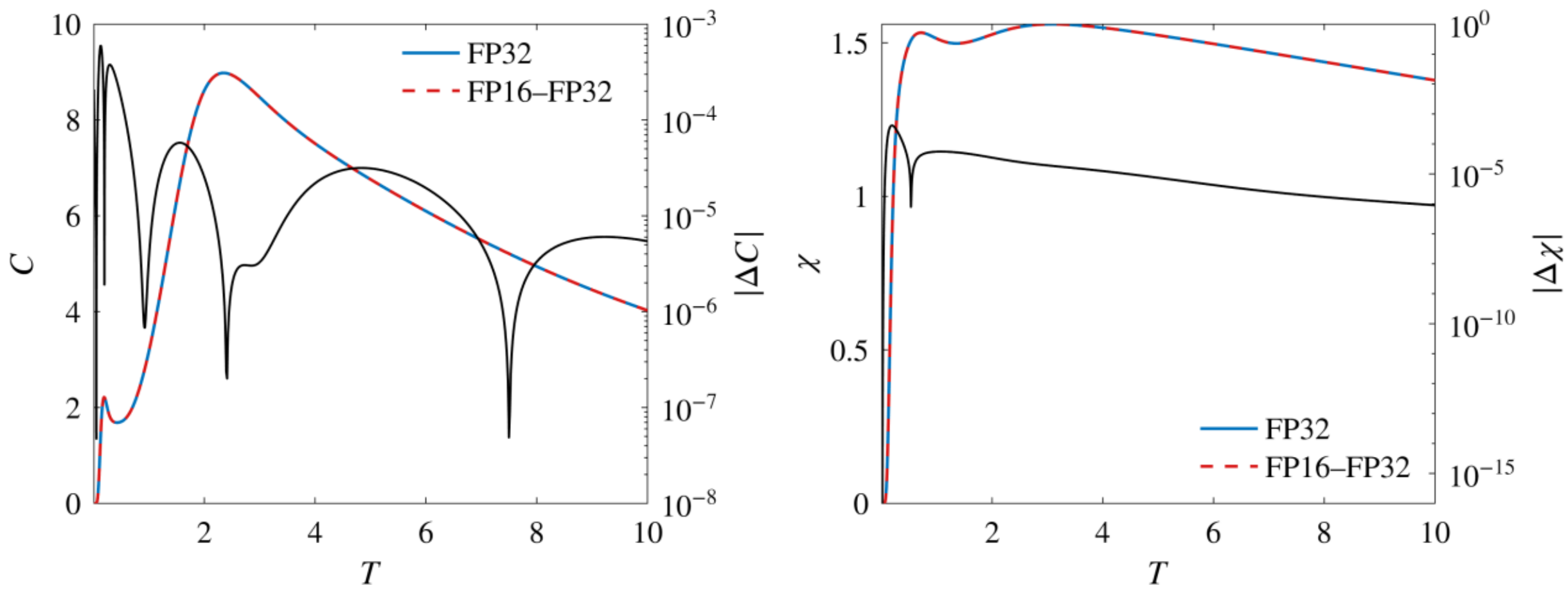

Figure 5: Same-seed comparison of FP32 and FP16 vector storage in symmetry-adapted FTLM for the $s = 3$ icosahedron ($R = 24$, $N_L = 60$, all $M$ sectors). Left: heat capacity; right: zero-field susceptibility $\chi$ ($g = 1$). The curves coincide at the resolution of the plot. The pointwise deviation, $|\Delta C|$ or $|\Delta\chi|$, is plotted against the logarithmic right axis.

**4.3 Benchmarks.** Table 1 summarizes the benchmark systems. As a first workstation-scale example, consider the $M = 0$ sector of the $s = 1/2$ icosidodecahedron ($R = 24$ real Gaussian starting vectors per block, $N_L = 60$ Lanczos steps, batch width $B = 8$, bitmap lookup). The ten per-irrep blocks (Figure 2) complete in 2–16 $s$ each, and the full ten-irrep sweep (twenty parity-resolved blocks, Section 2.2) takes 96.5 $s$ end to end within 2.3 GB of device memory, dominated by the Lanczos iterations.

**Table 1.** Benchmark systems. $\mathcal{D}_{\text{full}}$ is the full Hilbert-space dimension, $\mathcal{D}_0$ the dimension of the $M = 0$ sector.

| System | $N$ | $s$ | $N_B$ | Group | $\mathcal{D}_{\text{full}}$ | $\mathcal{D}_0$ |
|---|---|---|---|---|---|---|
| Icosidodecahedron | 30 | 1/2 | 60 | $I_h$ (120) | $2^{30} = 1.07 \times 10^9$ | 155 117 520 |
| Dodecahedron | 20 | 3/2 | 30 | $I_h$ (120) | $4^{20} = 1.10 \times 10^{12}$ | 86 981 744 944 |
| Square lattice 5×8 | 40 | 1/2 | 80 | $C_{2v} \ltimes T$ (160) | $2^{40} = 1.10 \times 10^{12}$ | 137 846 528 820 |
| Dodecahedron | 20 | 1 | 30 | $I_h$ (120) | $3^{20} = 3.49 \times 10^9$ | 377 379 369 |
| Icosahedron | 12 | 3 | 30 | $I_h$ (120) | $7^{12} = 1.38 \times 10^{10}$ | 786 588 243 |
| Kagome lattice | 36 | 1/2 | 72 | $C_{6v} \ltimes T$ (144) | $2^{36} = 6.87 \times 10^{10}$ | 9 075 135 300 |

The $M$-sector dimensions of the $s = 3/2$ dodecahedron ($8.7 \times 10^{10}$ at $M = 0$) exceed the full Hilbert space of the $s = 1/2$ icosidodecahedron by two orders of magnitude, but the permutation-symmetry reduction keeps the blocks within reach of a single B200 accelerator. All magnetization sectors are computed independently on the B200 node, with $R = 24$, $N_L = 60$, $B = 1$, and the ranking lookup.

**Table 2.** Per-irrep block dimension for the $s = 3/2$ dodecahedron at $M = 0$. Dimensions aggregate the $\Gamma^+/\Gamma^-$ sub-blocks of the spin-flip-extended run per parent irrep (Section 2.2).

| $\Gamma$ | $d_\Gamma$ | $\mathcal{D}_{M,\Gamma}$ |
|---|---|---|
| $A_g$ | 1 | 724 986 514 |
| $A_u$ | 1 | 724 739 466 |
| $T_{1g}$ | 3 | 2 174 409 448 |
| $T_{1u}$ | 3 | 2 174 648 732 |
| $T_{2g}$ | 3 | 2 174 409 448 |
| $T_{2u}$ | 3 | 2 174 648 732 |
| $F_g$ | 4 | 2 899 395 938 |
| $F_u$ | 4 | 2 899 388 178 |
| $H_g$ | 5 | 3 624 380 734 |
| $H_u$ | 5 | 3 624 125 950 |

The $M = 0$ sector completes in 4.8 h, including 45 minutes to build its entry table ($12.5 \times 10^9$ entries, 88 GB), which is sorted on node-local scratch, is then fully device-resident, and is reused across all irreps of the sector. The largest Lanczos vectors occur at $M = 1$ ($n_{\text{basis}} = 3.55 \times 10^9$; the $M = 0$ blocks of Table 2 are iterated as two $\Gamma\pm$ sub-blocks of roughly half the listed dimension). The full calculation over all magnetization sectors totals about 50 GPU-hours. From the $M = 0$ sector we obtain the ground-state energy $E_0 = -62.4388$ (FP32; deepening the recursion from 60 to 200 steps shifts the value by less than $10^{-4}$).

Figure 6 shows the FTLM heat capacity and susceptibility for the $s = 3/2$ dodecahedron over $T \in [0,5]$ obtained from Eqs. (5) and (7). The heat capacity has its maximum $C_{\max} = 9.61$ at $T = 1.74$, with a low-temperature shoulder near $T \approx 0.15$; the susceptibility maximum is $\chi_{\max} = 11.98$ at $T = 1.37$ ($g = 2$ convention).

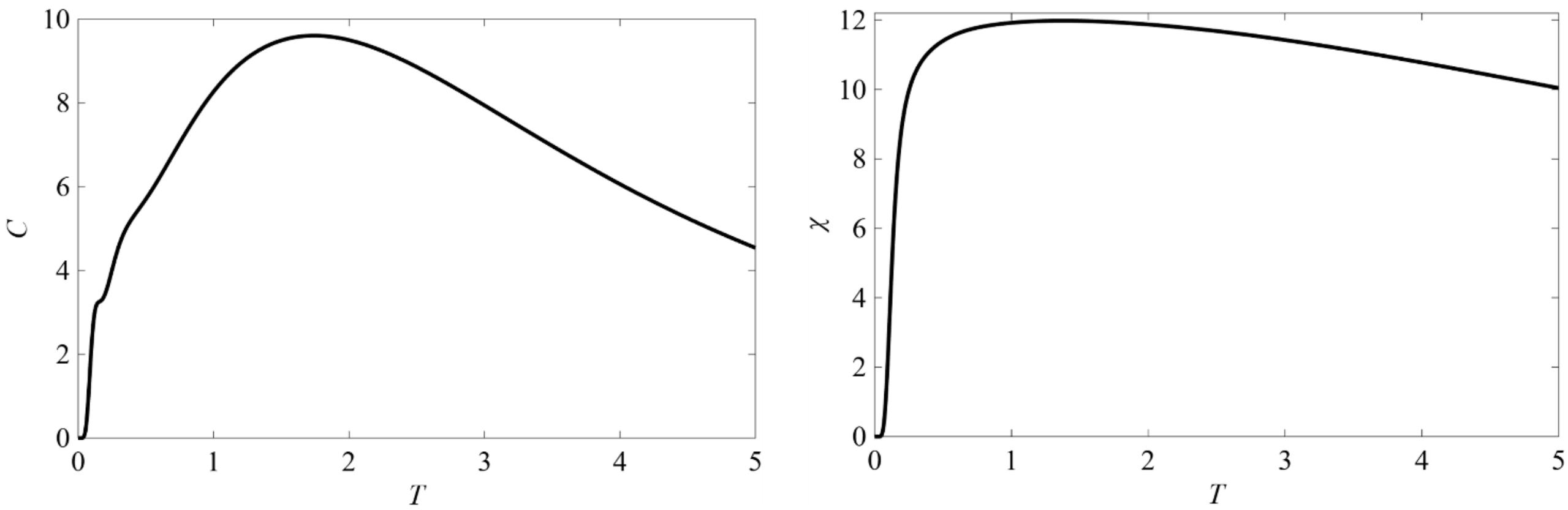

Figure 6. FTLM heat capacity $C$ (left) and zero-field susceptibility $\chi$ (right) of the $s = 3/2$ dodecahedron ($R = 24$, $N_L = 60$, all magnetization sectors; runs of Table 2). Units $k_B = \mu_B = 1$; the susceptibility axis uses $g = 2$, cf. Eq. (7).

As a test outside the polyhedral family, we consider the $s = 1/2$ Heisenberg antiferromagnet on the $5 \times 8$ square lattice with periodic boundary conditions, using its full space group (point group $C_{2v}$ combined with the $5 \times 8$ translations), $|G| = 160$, which the $M = 0$ spin flip extends to $|G| = 320$. Note that for this lattice, the full space group yields no block-dimension advantage over momentum-space techniques (Section 2.4). With $R = 24$, $N_L = 60$, $B = 1$, and the ranking lookup ($\mathcal{D}_{\text{full}} = 2^{40} > 2^{32}$), the $M = 0$ sector completes in 36 h using the default FP32 storage throughout. The 28 irreps of the space group ($\Sigma d_\Gamma{}^2 = 160$) split into 56 parity-resolved blocks at $M = 0$, with the largest blocks ($d_\Gamma = 4$) having dimension 1 723 070 831, corresponding to a 6.9 GB Lanczos vector. The entry table ($17.7 \times 10^9$ entries) is disk-sorted and then fully device-resident. Peak device memory is 149.1 GB, peak host memory 159.0 GB. The extremal Ritz value yields the ground-state energy $E_0 = -25.6505$ (independent chains at Krylov depth 200 agree to $10^{-6}$).

Finally, for a cross-platform comparison, Table 3 lists wall times of identical runs on the B200 node and on the workstation for four representative systems. In the FP32 default mode, the B200 completes the polyhedral benchmarks 4.7–5.3 times faster than the RTX 4000. For the kagome fragment the factor rises to 8.6× because the resident 11.6 GB entry table leaves the 20 GB card too little free memory for full batching. The half-precision mode shortens every run further on both cards, modestly on the fully resident polyhedra but by up to 3.0× on the workstation kagome run, where the halved vector footprint restores the full batch width. The Lanczos iteration dominates the wall time throughout. That the cross-platform factor stays far below the ratio of nominal memory bandwidths, reflects the random gather at the heart of the SpMV: its throughput is set by memory latency rather than bandwidth (hardware-counter profiles of the production kernel show the DRAM interface below 4% utilization at L2 hit rates near 90%) and batching several Lanczos chains through a single pass partly amortizes this latency.

Since the entry tables can be streamed from host or disk (Section 3), the Lanczos vectors are the binding memory constraint, and the wall time grows approximately linearly with the block dimension in the SpMV-dominated regime. The 191.5 GB of the B200 raise this ceiling by roughly an order of magnitude compared to the 20 GB workstation card. The half-precision storage mode halves this footprint and thereby roughly doubles the feasible block dimension.

**Table 3.** End-to-end wall times (in seconds) for identical FTLM runs ($R = 24$, $N_L = 60$, all $M \geq 0$ sectors; kagome: $M = 0$ sector only, $R = 8$) on the B200 node and on the RTX 4000 workstation. Columns labeled FP32 use the default vector storage; columns labeled FP16 use the half-precision storage mode of Section 4.2. The speedup column compares the FP32 runs.

| System | B200 (FP32) | B200 (FP16) | RTX (FP32) | RTX (FP16) | Speedup |
|---|---|---|---|---|---|
| Icosidodecahedron, $s = 1/2$ | 91.5 | 81.1 | 433.4 | 387.8 | 4.74× |
| Dodecahedron, $s = 1$ | 262.2 | 231.4 | 1313.6 | 1210.5 | 5.01× |
| Icosahedron, $s = 3$ | 1332.4 | 1136.6 | 7011.2 | 6579.0 | 5.26× |
| Kagome $N = 36$, $s = 1/2$ | 1013.4 | 665.0 | 8712.7 | 2944.7 | 8.60× |

# 5. Summary and outlook

We have extended matrix-free GPU-accelerated FTLM to the full permutation symmetry of Heisenberg spin clusters, including non-Abelian point and space groups. Central elements are the state-local matrix-element formula of Eq. (15), evaluated inside the SpMV, and a kernel parallelization over output-row pairs that remains efficient for all irrep dimensions. On a single

B200 accelerator, this brings system sizes into reach that were previously associated with distributed-memory CPU installations: the complete thermodynamics of the $s = 3/2$ dodecahedron was computed in ~50 hours with half-precision vector storage.

The implementation is restricted to the isotropic Heisenberg Hamiltonian. Anisotropic exchange or single-ion terms would modify the bond term of Eq. (3), and models that do not conserve $M$ would require deeper changes. A natural extension concerns spatially symmetric clusters whose anisotropic interactions are compatible with the spatial symmetry. Such models retain symmetries that combine spin rotations with site permutations (for an icosahedron with bond-directed symmetric anisotropic exchange, the resulting group is still isomorphic to $I_h$). We note, however, that a general spin-space rotation maps an $S^z$ product state onto a linear combination of many product states, so the state-local orbit machinery of Section 2.3 no longer applies as is; only spin-space operations that permute the product basis, such as the discrete flip used here, preserve this structure. A symmetry-adapted matrix-free FTLM for such models therefore requires new construction steps and lies beyond the scope of the present work.

The natural next step is toward still larger systems, and it is instructive to ask where the single-accelerator envelope ends (cf. Figure 1). The $N = 42$ kagome torus, the largest spin-1/2 FTLM computation reported to date [34], is a useful marker. Its 14-cell geometry admits no space group of order larger than 112 (including the $M = 0$ spin flip), so the $M = 0$ sector of dimension $5.4 \times 10^{11}$ comprises about $4.8 \times 10^{9}$ orbits. A single-partner-row block of an irrep of dimension $d_\Gamma$ collects roughly $d_\Gamma$ times that number of basis states (Eq. (13)), and the actual block table requires the irreps of the 112-element group. With half-precision storage for the Lanczos vectors, streamed entry tables, and the representative indexing widened to 64 bits, reaching this system on a single B200-class node appears feasible at least for the lower irrep dimensions, but the largest blocks approach the device limit. Two further directions follow naturally. Distributing the Lanczos vectors, together with the corresponding slices of the basis, across several GPUs would remove the memory ceiling altogether, in analogy to distributed-memory CPU implementations, and would carry GPU-based FTLM to system sizes that approach or exceed what has been reached with massively parallel CPU installations.

**Acknowledgments.** S.G.T. was supported by the Deutsche Forschungsgemeinschaft (DFG) under Project 535298924.

**Code and data availability.** The complete source code, including the MATLAB driver, CUDA MEX kernels is available at https://github.com/ghasdeke/ftlm-pg-gpu under the Apache-2.0 license, archived at DOI: 10.5281/zenodo.21445872.

## Appendix A. Derivation of the matrix-element formula (15)

The transfer operators of Section 2.3 satisfy the matrix-unit relations $(P^{\Gamma}_{\alpha,\beta})^{\dagger} = P^{\Gamma}_{\beta,\alpha}$ and $P^{\Gamma}_{\alpha,\beta}\, P^{\Gamma'}_{\gamma,\delta} = \delta_{\Gamma\Gamma'}\, \delta_{\beta,\gamma}\, P^{\Gamma}_{\alpha,\delta}$, which follow from the group-orthogonality theorem. Re-indexing the group sum in the definition of $P^{\Gamma}_{\alpha,\beta}$ ($g \to g\, g_e^{-1}$) for a fixed group element $g_e$ gives the transformation rule

$$P^{\Gamma}_{\alpha,\beta}\, U(g_e) = \sum_{\gamma} [\boldsymbol{\rho}_{\Gamma}(g_e)]_{\beta,\gamma}\, P^{\Gamma}_{\alpha,\gamma} \quad . \text{ (A.1)}$$

A coefficient vector attached to the second index of $P$ therefore contracts with the matrix elements of $\boldsymbol{\rho}_{\Gamma}(g_e)$ in transposed arrangement, which is the origin of the transpose in Eq. (15).

Consider one off-diagonal term of $H$ that maps the representative state $|r\rangle$ to $c_a|a\rangle$ with $|a\rangle = U(g_e)|r'\rangle$ and $r' = \min O(a)$ (Section 2.4). Since $H$ commutes with every $U(g)$, inserting the basis states of Eq. (11) gives for this contribution

$$M^{(e)}_{k',k} = c_a(\lambda_{r',k'}\, \lambda_{r,k})^{-1/2} \sum_{\beta',\beta} (V^{*}_{r'})_{\beta',k'} (V_r)_{\beta,k} \langle r'| P^{\Gamma}_{\beta',\beta}\, U(g_e)|r'\rangle \quad . \text{ (A.2)}$$

Applying Eq. (A.1) inside the bracket and recognizing the Gram matrix of Eq. (10), $\langle r'|P^{\Gamma}_{\beta',\gamma}|r'\rangle = M_{\Gamma}(r')_{\beta',\gamma} = \sum_{k''} \lambda_{r',k''} (V_{r'})_{\beta',k''} (V^{*}_{r'})_{\gamma,k''}$, the $\beta'$-sum collapses by the orthonormality of the eigenvector columns of $V_{r'}$, leaving

$$M^{(e)}_{k',k} = c_a(\lambda_{r',k'}/\lambda_{r,k})^{1/2} [V^{\dagger}_{r'}\, \boldsymbol{\rho}_{\Gamma}(g_e)^{T}\, V_r]_{k',k} \quad , \text{ (A.3)}$$

which is Eq. (15). The choice of $g_e$ is free up to the stabilizer of $r'$: replacing $g_e \to g_e h$ with $h \in \mathrm{Stab}(r')$ multiplies the contraction by $\boldsymbol{\rho}_{\Gamma}(h)^{T}$; on the range of the Gram matrix one has $\boldsymbol{\rho}_{\Gamma}(h)^{*}\, V_{r'} = V_{r'}$ for every $h \in \mathrm{Stab}(r')$ (Section 2.3), hence $V^{\dagger}_{r'}\, \boldsymbol{\rho}_{\Gamma}(h)^{T} = V^{\dagger}_{r'}$ and Eq. (A.3) is unchanged. Site permutations act by $U(g)|n\rangle = |g \cdot n\rangle$ and representation matrices compose as $\rho_\Gamma(g_1 g_2) = \rho_\Gamma(g_1)\rho_\Gamma(g_2)$. With these conventions the released code evaluates Eq. (A.3) directly. Finally, Eq. (A.3) is the contribution of a single spin-flip entry. The full matrix element between two representatives is the sum over all entries connecting them, which the SpMV accumulates term by term (Section 3).